\documentclass[12pt,oneside,reqno]{amsart}
\usepackage{amssymb, amsmath, amsfonts, amsthm, graphicx, subfig, color,soul,array,booktabs,lineno,prettyref,cite,hyperref}

\begin{document}
\title{A Method for Massively Parallel Analysis of Time Series}
\author{Yi H. Yan}
\address{Yi H. Yan: University of Georgia, Institute of Bioinformatics, Athens, 30602, USA}
\author{Elizabeth D. Trippe}
\address{Elizabeth D. Trippe: University of Georgia, Department of Mathematics, Athens, 30602, USA,  Middle Georgia State University, Department of Mathematics, Macon, 31206, USA}
\author{Juan B. Gutierrez$^*$}
\address{Juan B. Gutierrez: \href{mailto:jgutierr@uga.edu}{jgutierr@uga.edu} University of Georgia, Department of Mathematics and Institute of Bioinformatics, Athens, 30602, USA}
	\date{\today}
\begin{abstract}						
			Quantification of system-wide perturbations from time series -omic data (i.e. a large number of variables with multiple measures in time) provides the basis for many down-stream hypothesis generating tools. Here we propose a method, Massively Parallel Analysis of Time Series (MPATS) that can be applied to quantify transcriptome-wide perturbations. The proposed method characterizes each individual time series through its $\ell_1$ distance to every other time series. Application of MPATS to compare biological conditions produces a ranked list of time series based on their magnitude of differences in their $\ell_1$ representation, which then can be further interpreted through enrichment analysis. The performance of MPATS was validated through its application to a study of IFN$\alpha$ dendritic cell responses to viral and bacterial infection. In conjunction with Gene Set Enrichment Analysis (GSEA), MPATS produced consistently identified signature gene sets of anti-bacterial and anti-viral response. Traditional methods such as EDGE and GSEA Time Series (GSEA-TS) failed to identify the relevant signature gene sets. Furthermore, the results of MPATS highlighted the crucial functional difference between STAT1/STAT2 during anti-viral and anti-bacterial response. In our simulation study, MPATS exhibited acceptable performance with small group size (n = 3), when the appropriate effect size is considered. This method can be easily adopted for other -omic data types. 
\end{abstract}
\date{\today}

\maketitle 


\section{Introduction}
The advent of high throughput molecular technologies has allowed us to measure a large number of variables simultaneously. Transcriptomic experiments typically measure the abundance of 20,000 or so transcripts. The number of variables measured by other ``omic'' technologies such as lipidomic, proteomic, glycomic and metabolomic studies are of similar magnitude. Time series -omic data refers to molecular snapshots taken using these -omic technologies on a time trajectory. Time series -omic data captures time dependent molecular dynamics and is a powerful tool in the study of disease progression \cite{becavin2011improving}, developmental process \cite{roy2010identification} and vaccination \cite{palermo2011genomic}.

 A variety of tools have been developed for the analysis of time series -omic data. Methods such as MESS\cite{berk2012longitudinal} and EDGE\cite{storey2005significance} aim to discover individual gene expression time series that are significantly different between two experimental conditions. Other methods such as TcGSA \cite{hejblum2015time} ,CAMERA \cite{wu2012camera} , and GSEA-TS aim to find time series of pre-defined gene sets that are significantly different among groups. Furthermore, clustering based time series analysis tools \cite{straube2015linear,ernst2005clustering} have also been developed. 
 
Current time series analysis tools focus on finding gene expression time series or pre-defined gene sets that most likely have changed between groups, but ignore the changes in pairwise gene dynamics. Pair wise dynamics between genes can be quantified using correlation, mutual information or a distance metric. Differential correlation has been used to study gene association with the clinical outcome of lung cancer \cite{shedden2005differential} and estrogen receptor modulation in hormonal cancers \cite{hsiao2016differential}. In this paper, we propose a novel method, Massive Parallel Analysis of Time Series (MPATS), for the analysis of time series data based on the detection of differential pairwise $\ell_1$ distances. MPATS is based on the use of a linear mixed model for gene expression time series. By characterizing each gene expression time series through its $\ell_1$ distance to all other time series within the system, MPATS allows for the clear quantification of the impact of perturbation on each time series in the context of the biological system.

A simulation study was conducted to demonstrate the statistical performance of our method and a motivational study was used to validate the biological relevance of the analysis result produced by MPATS. In the motivational study, time series transcriptomic data were used to characterize the responses of IFN$\alpha$ dendritic cells to different antigens \cite{banchereau2014transcriptional}. In contrast with traditional time series analysis methods such as EDGE and GSEA-TS, MPATS was able to identify signature gene modules that distinguish between anti-viral and anti-bacterial responses of IFN$\alpha$ dendritic cells. 


This paper is organized as follows: section 2 contains the derivations, theoretical justifications and description of our method.  Section 3 provides the assessment of statistical power of our method based on simulated data. Section 4 contains the application of our method to a recent study of IFN$\alpha$ dendritic cells. Section 5 provides discussion of the performance of our method and future directions. 

 
\section{Method Description} 



\subsection{Linear Mixed Model of Time Series}

	Each individual time series is denoted $G_{ij}(t)$ and represents the abundance of gene $i$ from individual $j$ at time point $t$, where $i \in \{1,2,...n\}$, $j \in \{1,2,...m\}$ and $t \in \{1,2 ... h\}$. We used a mixed linear model \cite{storey2005significance} for the time series where:
	\begin{equation}\label{eq:1}
		G_{ij}(t) = g_{ij}(t) + \nu_t + \epsilon.
	\end{equation}
	The above model implies that the observed entity $G_{ij}(t)$ can be explained by the additive effect of its time dependent mean response $g_{ij}(t)$, individual and time based variation $\nu_t$, and instrumentation error $\epsilon$. We assume that both $\nu_t$ and $\epsilon$ follow Gaussian distributions, $\nu_t \sim  \mathcal{N}(0,\nu_t)$ and $\epsilon \sim  \mathcal{N} (0,\epsilon)$. Equation \prettyref{eq:1} can be further simplified to: 
	\begin{equation}
	G_{ij}(t) \sim \mathcal{N} \left(g_{ij}(t),\nu_t + \epsilon \right).
	\end{equation}

\subsection{$\ell_1$ Distance Between Two Time Series}	
	Let $\ell_1 (p,q,j)$ denote the $\ell_1$ distance between the mean response curves of genes $p$ and $q$ for subject $j$. 
	\begin{equation} \label{eq:3}
		\ell_1(p,q,j) = \sum_{t = 1}^h |g_{pj}(t) - g_{qj}(t)|.
	\end{equation}
	Let $X(p,q,j)$ be the observed $\ell_1$ distance between the two genes $p$ and $q$ for subject $j$.
	\begin{equation}
		X(p,q,j) = \sum_{t = 1}^h x(p,q,j,t),
	\end{equation}
	where 
	\begin{equation}
	x(p,q,j,t) = |G_{pj}(t) - G_{qj}(t)| \sim |\mathcal{N} \left(\mu_t,\sigma_t^2\right)|,
	\end{equation}
	and
	\begin{equation} \label{eq:6}
      \begin{split}
        \mu_t &= g_{pj}(t) - g_{qj}(t),\\
        \sigma_t^2 &= 2(\nu_t + \epsilon)
      \end{split}
	\end{equation}
	Through numerical investigation, Tsagris et al \cite{tsagris2014folded} saw that for any folded normal distribution $F$, where 
    \begin{equation}
    	F \sim |\mathcal{N}(\mu,\sigma^2)|,
    \end{equation}
	if $\mu \geq 3 \sigma^2$, then $F$ can be well approximated with a normal distribution with mean $|\mu|$ and variance $\sigma^2$. 

	Assuming that for any $t$, $|\mu_t| \geq 3 \sigma_t^2$, then 
	\begin{equation}
		X(p,q,j)\sim \sum_{t = 1}^h \mathcal{N}(\mu_t,\sigma_t^2).
	\end{equation}
	Which means
		\begin{equation}
		X(p,q,j) \sim \mathcal{N}(\sum_{t = 1}^h |\mu_t|, \sum_{t = 1}^h \sigma_t^2).
		\end{equation}
	According to equation \prettyref{eq:6} and \prettyref{eq:3} $\sum_{t = 1}^h |\mu_t| = \ell_1(p,q,j)$, and 
	\begin{equation} 
			X(p,q,j) \sim \mathcal{N}(\ell_1(p,q,j), \sum_{t = 1}^h \sigma_t^2).
	\end{equation} 
	Which implies that given sufficiently small variance for all $t$, the observed $\ell_1$ distance between two gene expression time series follows an approximate normal distribution with mean equal to the true $\ell_1$ distance.
	
	In the case that $|\mu_t| < 3 \sigma_t^2$, the mean of $X(p,q,j)$ does not necessarily equal to $\ell_1(p,q,j)$. 
	
	Let $c$ be a constant, such that 
	\begin{equation}
	c > \max 10 \sigma_t^2 \quad \forall \quad t \in \{1,2,3 ...h\}.
	\end{equation} 
	Then let
	\begin{equation}
        \begin{split}
         y_1(p,q,t,c,j) &= G_{pj}(t) - G_{qj}(t) + c,\\
          &\sim |\mathcal{N} \left(\mu_1,\sigma_t^2\right)|,
         \end{split}
	 \end{equation} 
     where 
     \begin{equation}
        \mu_1 = g_{pj}(t) - g_{qj}(t) + c. 
     \end{equation}
     Let
	\begin{equation}
	 	\begin{split}
	 	y_2(p,q,t,c,j) &= G_{qj}(t) - G_{pj}(t) + c,\\
	 	&\sim |\mathcal{N} \left(\mu_2,\sigma_t^2\right)|,
	 	\end{split}
	\end{equation} 
    where 
    \begin{equation}
        \mu_2 = g_{qj}(t) - g_{pj}(t) + c.
    \end{equation}
	Because $c > \max 10 \sigma_t^2 \quad \forall \quad t \in \{1,2,3 ...h\}$,
	\begin{equation}
		\begin{split}
		y_1(p,q,t,c,j) &\sim \mathcal{N}(\mu_1,\sigma_t^2),\\
		y_2(p,q,t,c,j) &\sim \mathcal{N}(\mu_2,\sigma_t^2).\\
		\end{split}
	\end{equation}
	Furthermore, 
	\begin{equation}
	  \max(\mu_1,\mu_2) = |g_{pj}(t) - g_{qj}(t)| + c. 
	\end{equation}
	Let 
	\begin{equation}
		z(p,q,t,c,j) = 
		\begin{cases} 
		y_1(p,q,t,c,j) &\mbox{if } \mu_1 \geq \mu_2, \\ 
		y_2(p,q,t,c,j) &\mbox{if } \mu_1 < \mu_2.
		\end{cases} 
	\end{equation} 
	Let
	\begin{equation} 
		Z(p,q,c,j) = \sum_{t=1}^{h} z(p,q,t,c), 
	\end{equation}
	then, 
	\begin{equation} \label{eq:2}
				Z(p,q,c,j) \sim \mathcal{N}(\ell_1(p,q,j) + hc,\sum_{t = 1}^h \sigma_t^2). 
	\end{equation}	
	Which implies that $Z(p,q,c,j)$ follows a normal distribution with mean equal to the true $\ell_1$ distance plus some known constan t.

\subsection{Detection of Significantly Changed $\ell_{pq}$ Between Groups}

	Each individual gene expression time series can be characterized by its $\ell_1$ distances to every other gene expression time series. A perturbation of the system (e.g. disease progression) shifts the mean response curve of each gene expression time series with an unknown magnitude and direction. 
	
	The impact of the perturbation on the system can be quantified by the number of $\ell_1$ distances that have changed due to perturbation. The impact of the perturbation on each individual gene time series can be quantified by the number of pairwise distances to this gene that have changed.

In the case of a two-group experiment, we define two sets of subjects, where $S_1 = \{1,2,3,4 ... a\}$ and $S_2 = \{a+1,a+2 ... m\}$. For a given pair of entities $p$ and $q$, let $A_{pqc}= \{Z(p,q,j,c) : j \in S_1\}$ and $B_{pqc}= \{Z(p,q,j,c) : j \in S_2\}$. As shown in \prettyref{eq:2}, both $A_{pqc}$ and $B_{pqc}$ follow a normal distribution with mean corresponding to the $\ell_1$ distance between gene $p$ and $q$ within each group. Welch's test is used to determine if $A_{pqc}$ and $B_{pqc}$ have the same mean. 

Let $P(p,q)$ be equal to the p-value from Welch's test. The results of conducting Welch's test for every pair of $p$ and $q$ can be captured in a $n$ by $n$ matrix $M$, where 

$$
M_{pq} = 
\begin{cases}
1, P(pq) < \delta \\
0, P(pq) \geq \delta,
\end{cases}
$$

where $\delta $ is the threshold p value. 
	 
For a gene $p$, its contribution to the perturbation can be quantified by its perturbation score $(PS)$, where: 

\begin{equation}
 PS = \sum_{q=1}^{n}M_{pq} . 
\end{equation}

The genes are then ranked according to their perturbation score from highest to lowest. 

\section{Power Assessment Using Simulated Data}

\begin{figure}
		
	\centering
		\textbf{ROC Curve of Different Sample Size and Effect Size}\par\medskip
		\makebox[\textwidth][c]{\includegraphics[width= 1.3\textwidth]{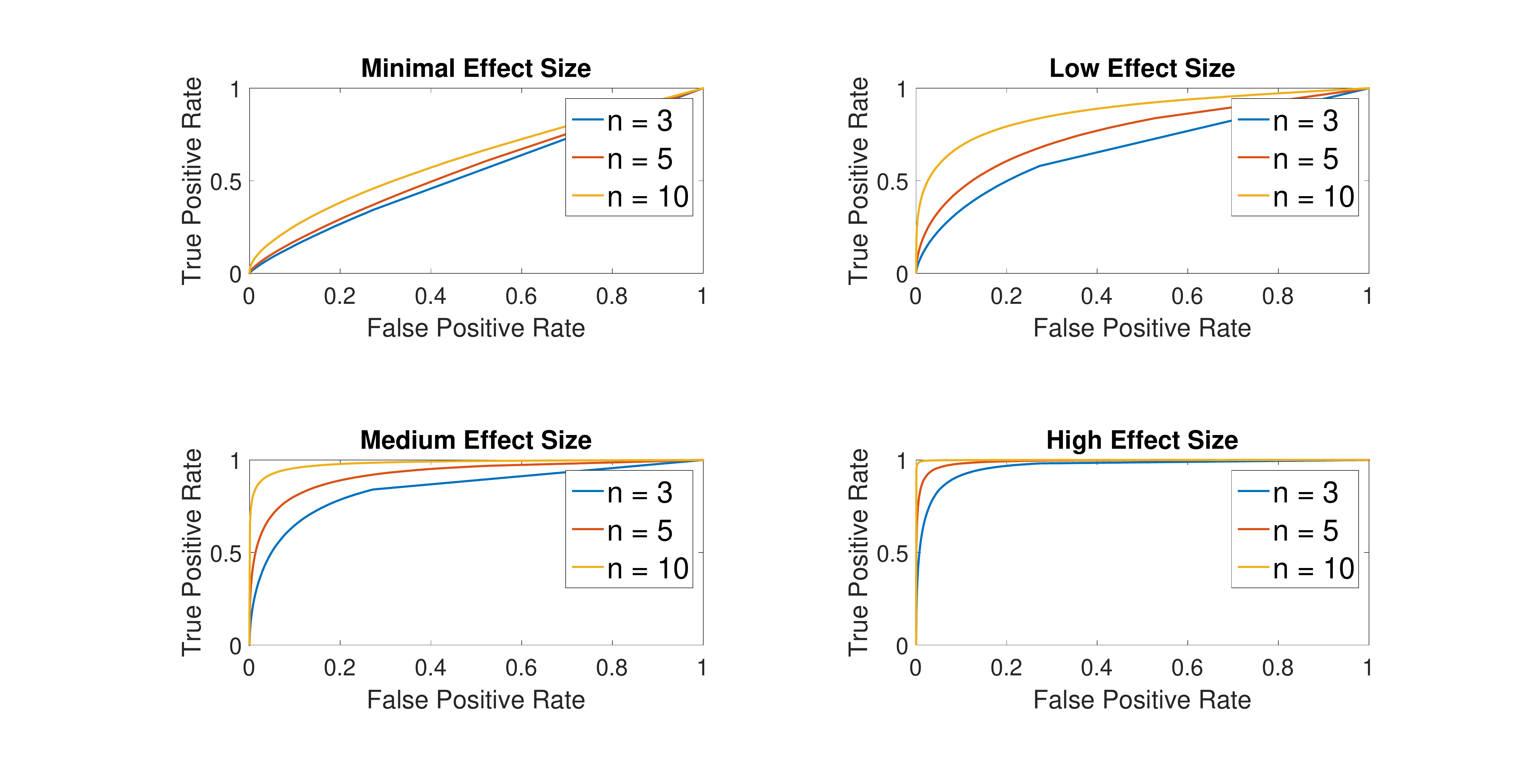}}
		
		\caption
		{ROC curves were generated for 4 different effect sizes. For each effect size, three different sample sizes were tested, $n = 3$, $n = 5$ and $n = 10$. For each simulation, 5000 gene expression time series were generated.}
		\label{Fig:1}
		
	\end{figure}

A simulation study was conducted to evaluate the statistical power of MPATS. The simulation framework was chosen based on the motivational study. For each simulation setup, 5000 gene expression time series were simulated and each time series contained 5 time points. The simulated values of each gene at each time point were drawn from a normal distribution with mean and variance estimated from the original data. 10 percent of the gene expression time series were perturbed in the treatment group. Receiver operating characteristic (ROC) curves were generated to explore the statistical performance of the proposed method for different group sizes and effect sizes. The effect size (ES) refers to the true difference in $\ell_1$ distance, and the category of effect sizes can be found in Table \ref{table1}. Statistical performance of the proposed method are presented in Figure \ref{Fig:1}. At a fixed specificity of $90\%$, MPATS has statistical power greater than $0.7$ with $n = 3$ for medium and high effect sizes. With $n =10$, MPATS can reach a power greater than $0.7$ with a fixed specificity of $90\%$ for low effect size. 

\begin{table}
	\centering
	\textbf{Effect Size Description}\par\medskip

	\begin{tabular}{rr}
		\addlinespace
		\toprule
		
		Category & $ES$ \\
		\midrule
		Minimal & $0 < ES \leq 0.25$\\
		Low & $0.25 < ES \leq 0.5$ \\
		Medium &$0.5 < ES \leq 1$\\
		High & $1 < ES$\\
		\bottomrule
	\end{tabular}
	\caption
	{Categories of Effect Size} \label{table1} 
\end{table}

\clearpage 

\section{Experimental Results}

\subsubsection{Motivating Study: IFN$\alpha$ dendritic cell response to antigen}

As part of a study to understand the different responses of dendritic cells to various antigens, microarray time series data were collected for IFN$\alpha$ dendritic cells and IL4 dendritic cells challenged with different antigens. Data were collected at  (1,2,6,12,24) hours after challenge. At each time point, three biological replicates were taken. The detailed description of the study can be found in Banchereau et al \cite{banchereau2014transcriptional}. 

For the purpose of the present paper, we focused on the time series of IFN$\alpha$ dendritic cells grown on media alone (MEDIA), challenged with H1N1 virus (H1N1) and heat killed \textit{Staphylococcus aureus} (HKSA). The data used in this study can be found on Gene Omnibus (GSE44720). 

\subsubsection{Application of MPATS, EDGE, and GSEA-Time Series}

MPATS and EDGE were used to conduct pairwise comparisons among the two treatment groups and the control group. GSEA-TS was applied to the time series data of the two treatment groups alone. During the application of MPATS, a q-value of $0.25$ was used as the threshold to determine whether a pairwise $\ell_1$ distance had changed significantly between groups. The ranked gene list produced by MPATS was analyzed using GSEA preranked \cite{subramanian2005gene}. The results of EDGE consisted of q-values for each gene. A ranked list was produced by ranking the genes according to their q-values in ascending order. The ranked gene list produced by EDGE was also analyzed using GSEA preranked. The gene sets tested for enrichment included: GSEA Hallmark gene sets, GO biological process gene sets, KEGG gene sets, and the signature gene modules described by Banchereau et al \cite{banchereau2014transcriptional}.   

\subsection{Clustering of pairwise $\ell_1$ distance representation of samples}

\begin{figure}
		
	\centering
  	\includegraphics[width=\textwidth]{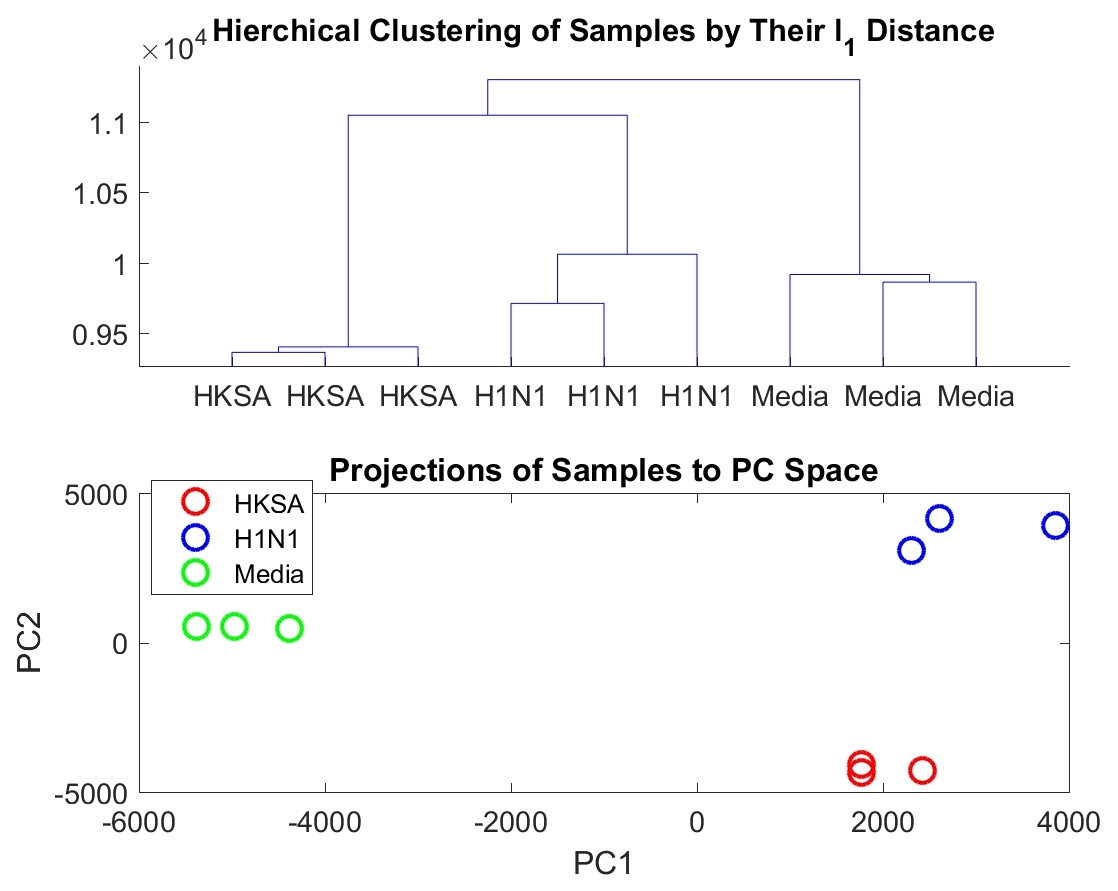}
    		\caption
		{Both hierarchical clustering of pairwise $\ell_1$ distance and projection into principle component space reveal underlying biological conditions.}
		\label{Fig:2}
		
	\end{figure}

\begin{figure}
		
\centering

		\includegraphics[width=\textwidth]{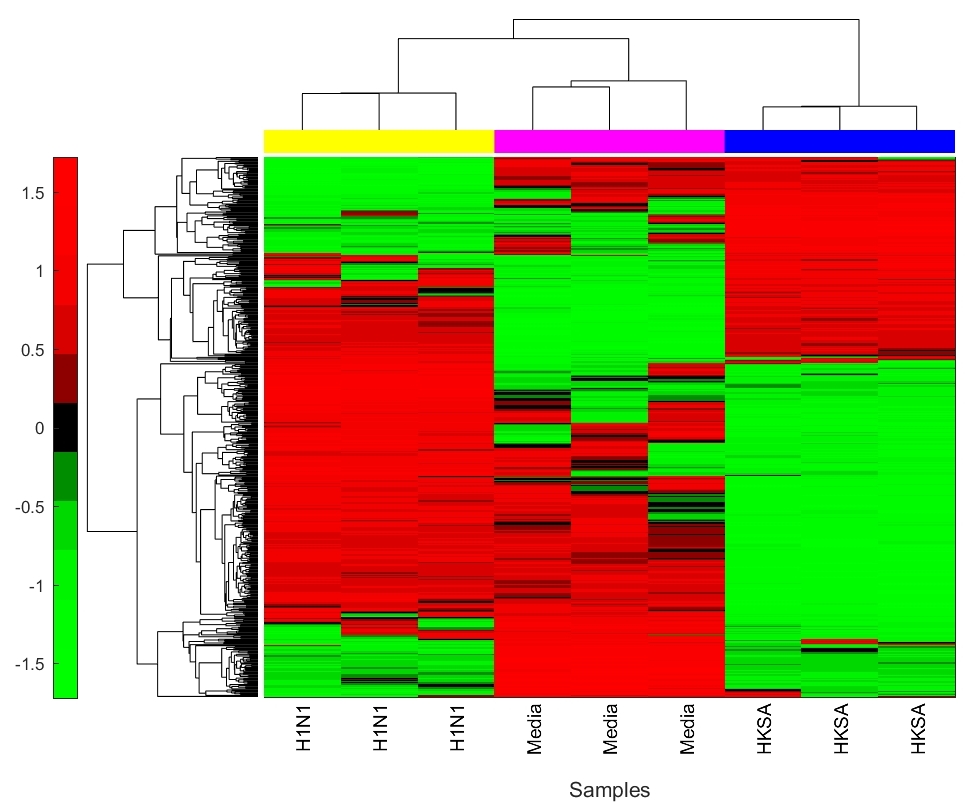}
		\caption
		{Top 100 $\ell_1$ distances with largest variance can be used to cluster the samples based on biological condition. Each row refers to the $\ell_1$ distance between two genes. Each column represents a sample.}
		\label{Fig:3}
		
	\end{figure}

Hierarchical clustering of the samples based on their pairwise $\ell_1$ distance representation reveals differences in biological conditions (Fig \ref{Fig:2}). Furthermore, projection of the data into the first and second principle component space also demonstrate clustering based on biological condition (Fig \ref{Fig:2}). The hierarchical clustering based on biological condition can be reproduced using only the top 100 $\ell_1$ distances with the largest variance across all three biological conditions (Fig \ref{Fig:3}). 

    
 \clearpage

	\begin{figure}
		
	\centering

		\includegraphics[width= \textwidth]{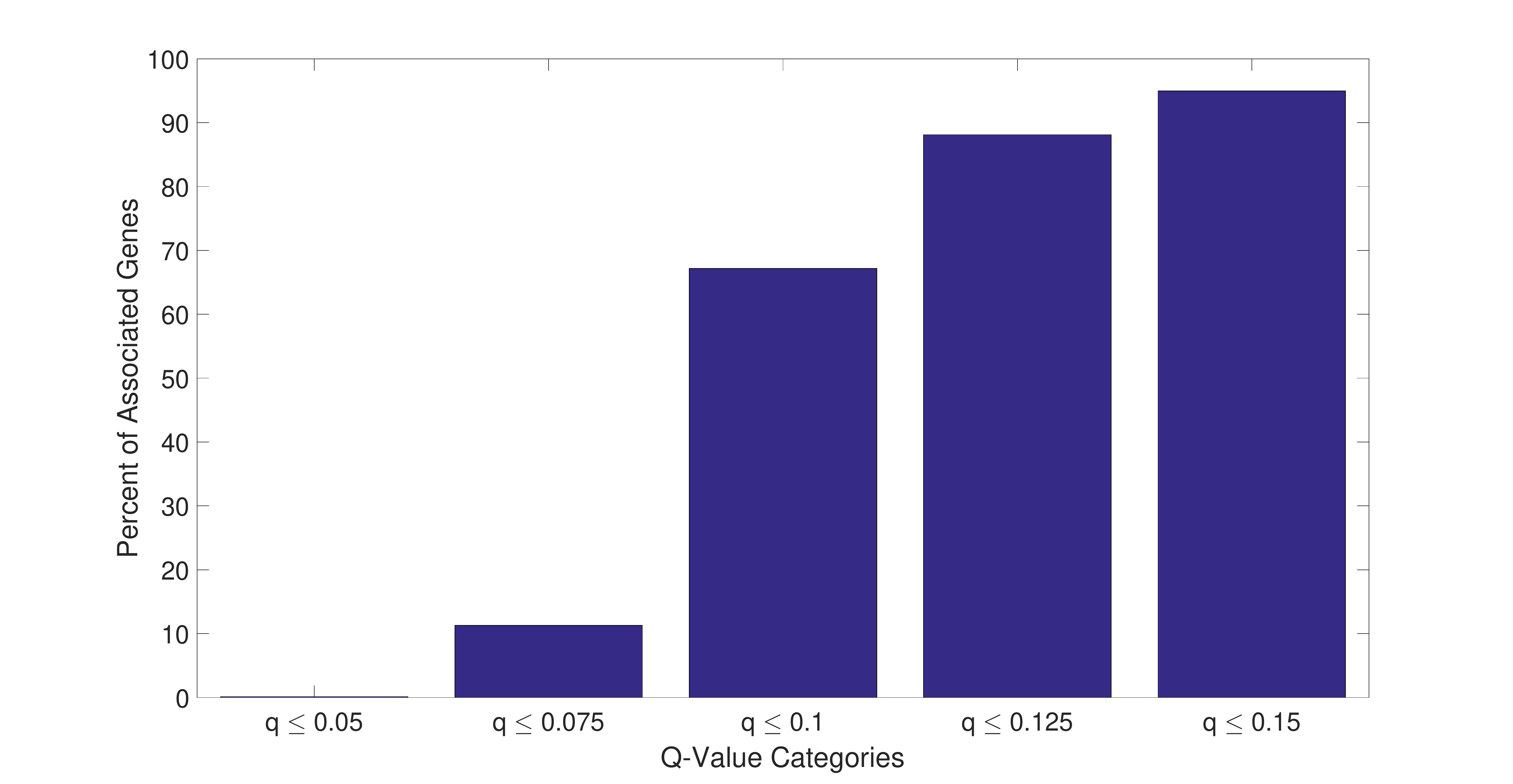}
		\caption
		{Percentage of genes associated with at least one $\ell_1$ distance smaller than a threshold (q).}
		\label{Fig:4}
		
	\end{figure}
   \clearpage

\begin{figure}
		
	\centering

		\includegraphics[width= \textwidth]{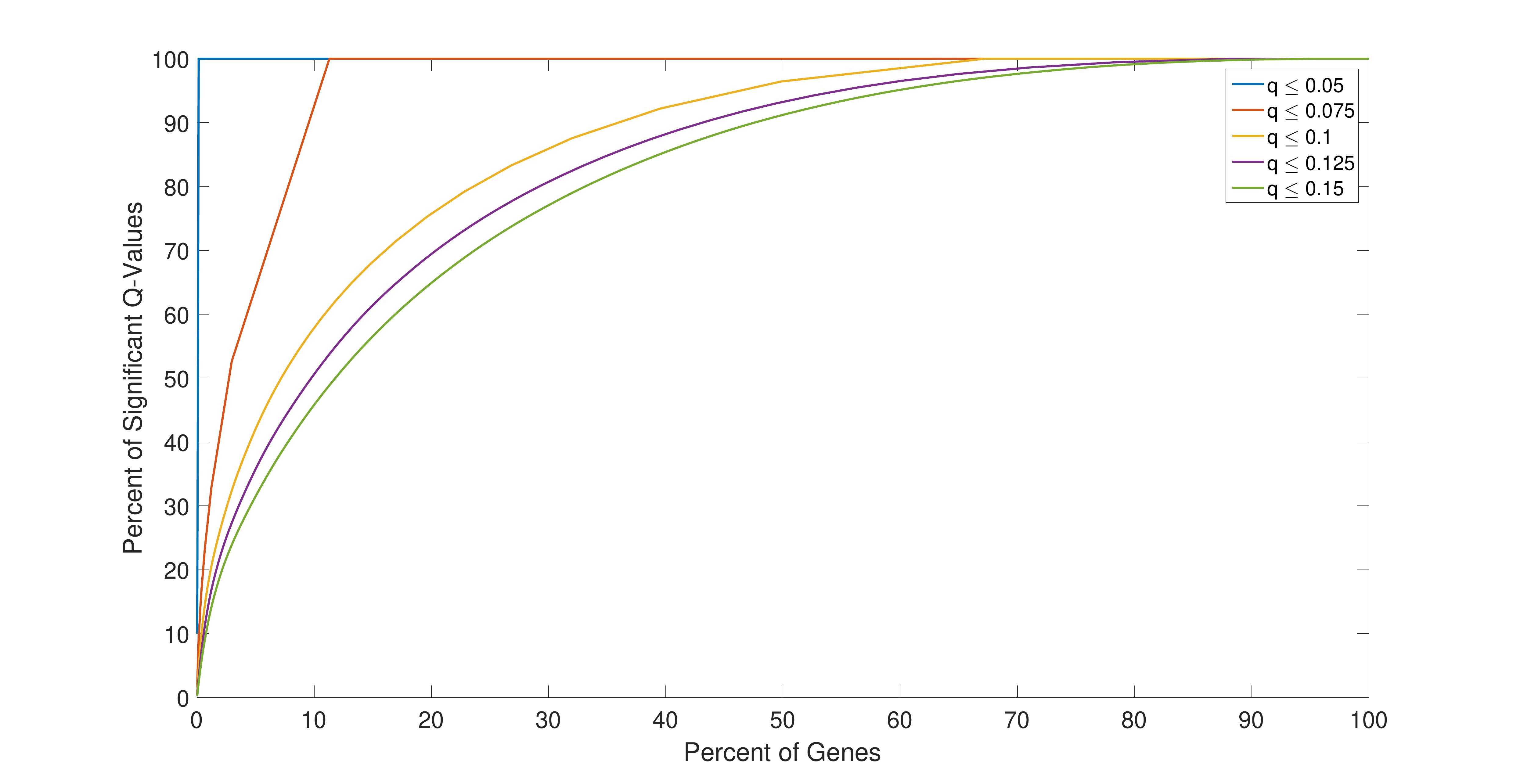}
		\caption
		{Highly significantly changed $\ell_1$ distance are associated with a small sub portion of the total genes.}
		\label{Fig:5}
		
	\end{figure}

\begin{figure}
    	\centering

		\includegraphics[width= \textwidth]{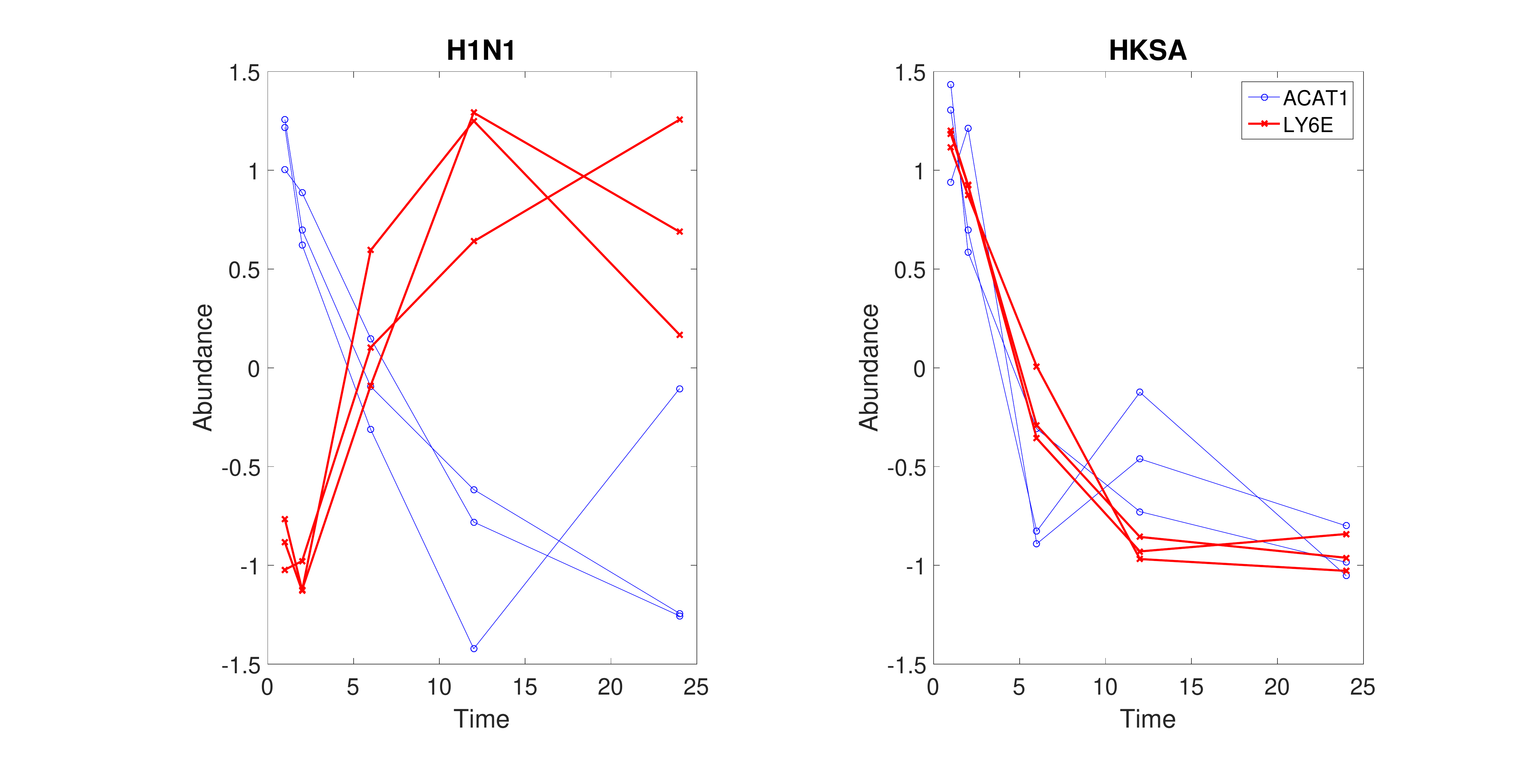}
		\caption
		{$\ell_1$ distance between ACAT1 and LYSE changed significantly $(q \leq 0.005)$ between the H1N1 challenged group and HKSA challenged group. Time series from all three biological replicated are displayed.}
		\label{Fig:6}
		
	\end{figure}

\begin{figure}
    	\centering

		\includegraphics[width= \textwidth]{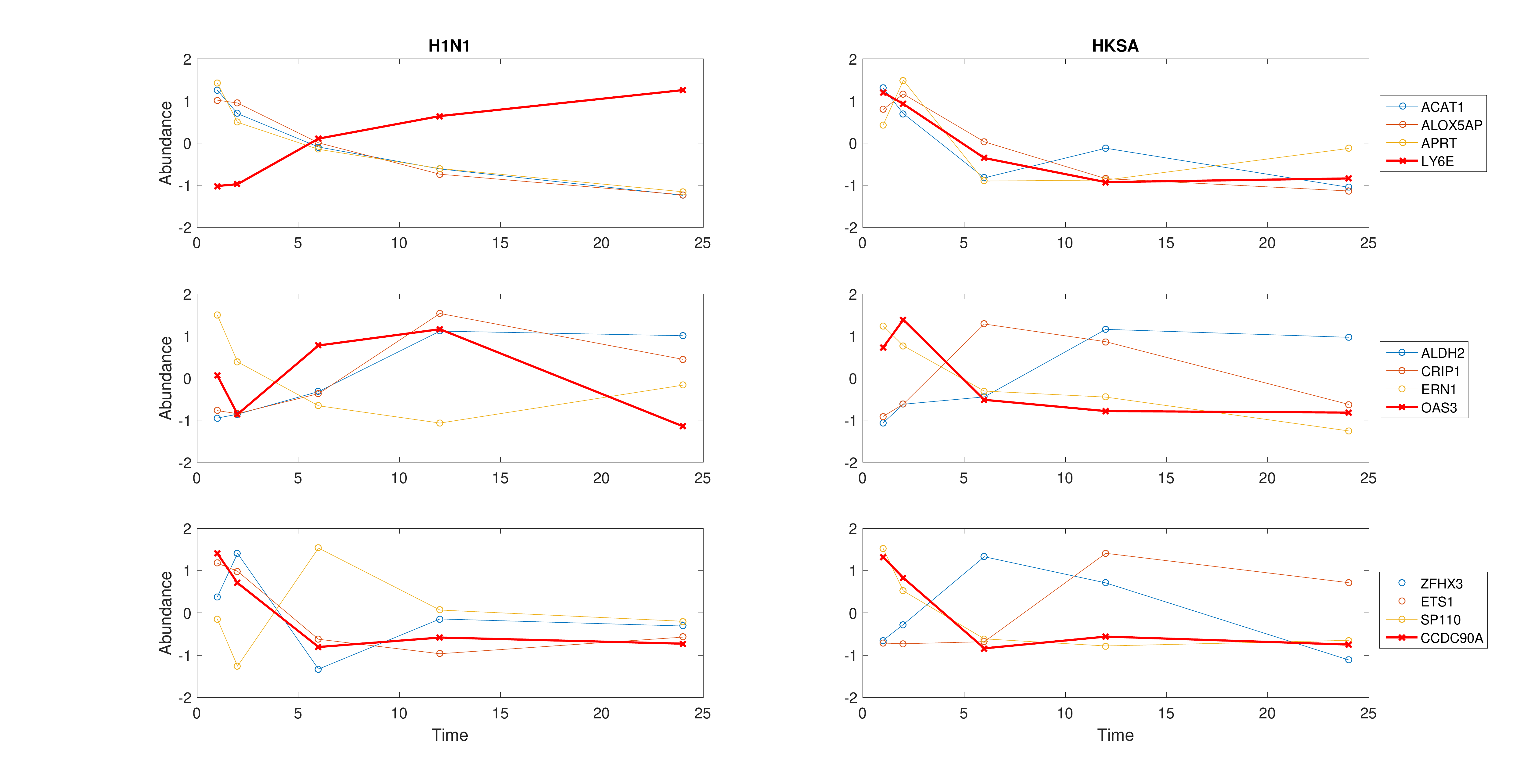}
		\caption
		{Examples of most significantly changed $\ell_1$ distances associated with genes of different ranking. The red solid lines are the time series expression profile of the gene of interest in each row. LY6E has rank 1, OAS3 has rank 100 and CCDC90A has rank 1000. The ranking of genes strongly associates with the changes in temporal pattern.}
		\label{Fig:7}
		
	\end{figure}

Only a subset of genes are associated with $\ell_1$ distances that have a small q-value. For the comparison between the H1N1 group and the MEDIA group, only $10\%$ of the genes are associated with a $\ell_1$ distance with a q-value smaller than $0.075$ (Figure \ref{Fig:4}). Furthermore, the highly significantly changed $\ell_1$ distances are associated with a small portion of the genes as shown in Figure \ref{Fig:5}. For the comparison the between the H1N1 group and the MEDIA group, $70\%$ of the $\ell_1$ distances with $q-value \leq 0.1$ are associated with $20\%$ of the genes. This non-uniform distribution of these significantly changed $\ell_1$ distances provides the basis for downstream enrichment analysis. An example of a significantly changed $\ell_1$ distance between two genes is shown in Figure \ref{Fig:6}. The ranking of each gene is based on the number of significantly changed $\ell_1$ distance associated with that gene. Figure \ref{Fig:7} shows that he ranking of each gene is strongly associated with the change in the time series of the gene itself.  
  
    \clearpage 
 
\subsubsection{MPATS Identifies Signature Gene Sets}

\begin{table}[htbp]
	\centering
	\textbf{Number of Perturbed Gene Sets Discovered by Each Method}\par\medskip
		\begin{tabular}{rrrrr}
		\addlinespace
		\toprule
		& HKSA    & H1N1  & H1N1\textbackslash SA \\
		\midrule
		MPATS & 326    & 348    & 270    \\
		EDGE  & 24    & 39     & 26    \\
		GSEA-TS & 48    & 42     & NA     \\
		\bottomrule
	\end{tabular}
	\caption{Three pairwise comparisons were conducted. Comparison between control group and treatment group challenged by SA (SA). Comparison between control group and treatment group challenged by H1N1 (H1N1). Comparison between the two treatment groups (H1N1\textbackslash SA).} 	\label{table2} 	
\end{table}

MPATS identified the greatest number of perturbed gene sets out of all three methods. The number of perturbed gene sets identified by each method are reported in Table \ref{table2}.The differences in the number of perturbed gene sets identified among the three methods highlight the sensitivity of MPATS. This increased sensitivity is due to the fact that MPATS ranks the genes based on their contribution to system-wide dynamic differences between biological conditions. The large number of gene sets perturbed highlights the tremendous impact of antigen challenge on the cell culture. A majority of gene sets and pathways identified are interferon response related pathways. The enrichment results are included in the supplementary material. 

In the original study, a list of signature modules were identified for INF$\alpha$ cell response to different antigens. Enrichment analysis of both MPATS and EDGE results captured 9 out of 10 signature modules of INF$\alpha$ response to H1N1 challenge, where GSEA-TS only captured 5. All three method performed worse for the analysis of signature response modules of anti-bacterial response, capturing only $30\%$ of the signature modules. This could be attributed to the low signal strength of these modules themselves in comparison to the signal strength of the H1N1 modules. 

\begin{table}[htbp]
	\centering
	\textbf{Number of Enriched Signature Modules}\par\medskip
		\begin{tabular}{rrrrr}
		\addlinespace
		\toprule
		& SA    & H1N1  \\
		\midrule
		MPATS & 11/30    & 9/10    \\
		EDGE  & 9/30    & 9/10     \\
		GSEA-TS & 12/30    & 5/10     \\
		\bottomrule
	\end{tabular}
	\caption{MPATS results are enriched in the signature modules of INF$\alpha$ cells challenged with different antigens.}	
\end{table}

\begin{figure}
		
	\centering

		\includegraphics[width= \textwidth]{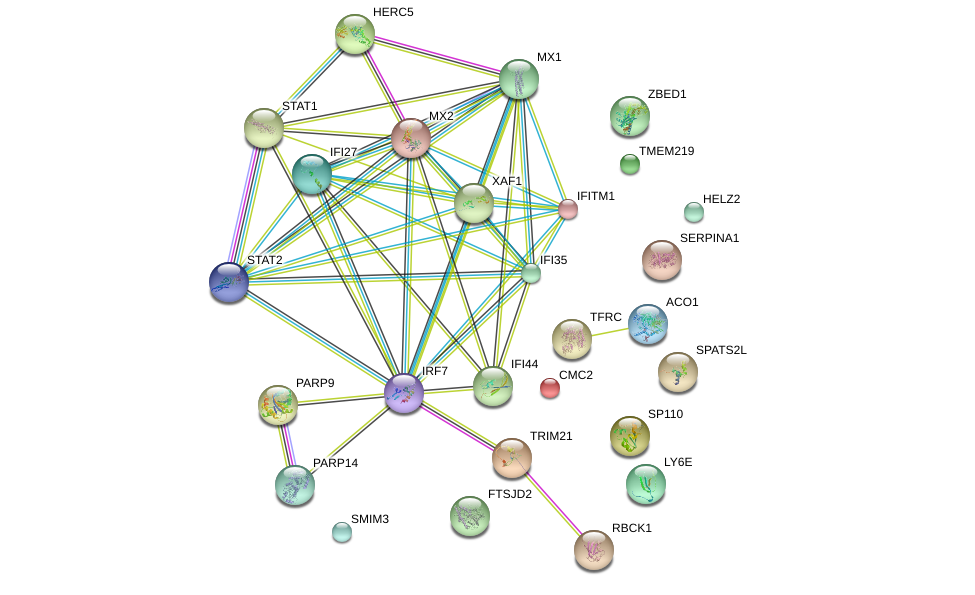}
		\caption
		{PPI Map of the top 30 genes for the comparison of H1N1 and HKSA.}
		\label{Fig:8}
		
	\end{figure}

The analysis comparing IFN$\alpha$ response to H1N1 and HKSA are enriched in interferon related gene sets, such as response to interferon gamma, TNF signal of beta kappa B and MTORC signaling. Analysis of the top 30 genes using String10 shows a enrichment in protein protein interaction (PPI) with $p-vlaue < 0.00001$. Furthermore, the PPIs are clustered around STAT1 and STAT2 (Figure \ref{Fig:8}). This result agrees with recent findings of the crucial differences in the role of STAT1/STAT2 in anti-viral and anti-bacterial responses \cite{blaszczyk2015stat2,hambleton2013stat2,hahm2005viruses,hofer2012mice}.

\section{Discussion}

In this paper, we presented a novel framework to quantify the perturbation of time series for two group experiments. The quantification of perturbation of time series provides context for downstream functional analysis. We were able to apply this framework to explore anti-viral and anti-bacterial response of IFN$\alpha$ dendritic cells. We found specific pathways that differentiate between viral and bacterial challenge of IFN$\alpha$ dendritic cells and identified STAT1/STAT2 as important regulatory elements differentiating these responses.

MPATS quantifies system wide perturbations and individual gene perturbations by characterizing each time series by its $\ell_1$ distance to every other time series. The biological significance of this intuitive method was demonstrated through analysis of the motivational study. The top ranked genes are not only enriched in PPIs but are also enriched in signature gene sets. The performance of MPATS remained stable through all three analyses, whereas EDGE and GSEA-TS identified varying numbers of perturbed gene sets. This is probably due to the fact that ranking by p-value of change is not biologically informative because a change of small magnitude with low variance can generate a low p-value. Ranking time series using variance normalized effect size should produce results similar to MPATS. The framework of MPATS is based on a linear mixed model and assumes normally distributed variance, and the use of pairwise dynamics to quantify magnitude of change for each individual entity can be easily expanded to other -omic data types.

In conclusion, MPATS complements existing time series analysis methods by providing a more biologically informative ranked list of genes of interest in addition to detecting genes that have experienced large perturbations but are overlooked by existing methods.
 
 \section*{Acknowledgments}
 Research reported in this publication was supported by a National Institute of Allergy and Infectious Diseases cooperative agreement of the National Institutes of Health under award number HHSN272201200031C.


\begin{thebibliography}{10}

\bibitem{banchereau2014transcriptional}
R.~Banchereau, N.~Baldwin, A.-M. Cepika, S.~Athale, Y.~Xue, I.~Y. Chun,
  P.~Metang, A.~Cheruku, I.~Berthier, I.~Gayet, et~al.
\newblock Transcriptional specialization of human dendritic cell subsets in
  response to microbial vaccines.
\newblock {\em Nature communications}, 5, 2014.

\bibitem{becavin2011improving}
C.~B{\'e}cavin, N.~Tchitchek, C.~Mintsa-Eya, A.~Lesne, and A.~Benecke.
\newblock Improving the efficiency of multidimensional scaling in the analysis
  of high-dimensional data using singular value decomposition.
\newblock {\em Bioinformatics}, 27(10):1413--1421, 2011.

\bibitem{berk2012longitudinal}
M.~Berk, C.~Hemingway, M.~Levin, and G.~Montana.
\newblock Longitudinal analysis of gene expression profiles using functional
  mixed-effects models.
\newblock In {\em Advanced Statistical Methods for the Analysis of Large
  Data-Sets}, pages 57--67. Springer, 2012.

\bibitem{blaszczyk2015stat2}
K.~Blaszczyk, A.~Olejnik, H.~Nowicka, L.~Ozgyin, Y.-L. Chen, S.~Chmielewski,
  K.~Kostyrko, J.~Wesoly, B.~L. Balint, C.-K. Lee, et~al.
\newblock Stat2/irf9 directs a prolonged isgf3-like transcriptional response
  and antiviral activity in the absence of stat1.
\newblock {\em Biochemical Journal}, 466(3):511--524, 2015.

\bibitem{ernst2005clustering}
J.~Ernst, G.~J. Nau, and Z.~Bar-Joseph.
\newblock Clustering short time series gene expression data.
\newblock {\em Bioinformatics}, 21(suppl 1):i159--i168, 2005.

\bibitem{hahm2005viruses}
B.~Hahm, M.~J. Trifilo, E.~I. Zuniga, and M.~B. Oldstone.
\newblock Viruses evade the immune system through type i interferon-mediated
  stat2-dependent, but stat1-independent, signaling.
\newblock {\em Immunity}, 22(2):247--257, 2005.

\bibitem{hambleton2013stat2}
S.~Hambleton, S.~Goodbourn, D.~F. Young, P.~Dickinson, S.~M. Mohamad,
  M.~Valappil, N.~McGovern, A.~J. Cant, S.~J. Hackett, P.~Ghazal, et~al.
\newblock Stat2 deficiency and susceptibility to viral illness in humans.
\newblock {\em Proceedings of the National Academy of Sciences},
  110(8):3053--3058, 2013.

\bibitem{hejblum2015time}
B.~P. Hejblum, J.~Skinner, and R.~Thi{\'e}baut.
\newblock Time-course gene set analysis for longitudinal gene expression data.
\newblock {\em PLoS Comput Biol}, 11(6):e1004310, 2015.

\bibitem{hofer2012mice}
M.~J. Hofer, W.~Li, P.~Manders, R.~Terry, S.~L. Lim, N.~J. King, and I.~L.
  Campbell.
\newblock Mice deficient in stat1 but not stat2 or irf9 develop a lethal cd4+
  t-cell-mediated disease following infection with lymphocytic choriomeningitis
  virus.
\newblock {\em Journal of virology}, 86(12):6932--6946, 2012.

\bibitem{hsiao2016differential}
T.-H. Hsiao, Y.-C. Chiu, P.-Y. Hsu, T.-P. Lu, L.-C. Lai, M.-H. Tsai, T.~H.-M.
  Huang, E.~Y. Chuang, and Y.~Chen.
\newblock Differential network analysis reveals the genome-wide landscape of
  estrogen receptor modulation in hormonal cancers.
\newblock {\em Scientific reports}, 6, 2016.

\bibitem{palermo2011genomic}
R.~E. Palermo, L.~J. Patterson, L.~D. Aicher, M.~J. Korth, M.~Robert-Guroff,
  and M.~G. Katze.
\newblock Genomic analysis reveals pre-and postchallenge differences in a
  rhesus macaque aids vaccine trial: insights into mechanisms of vaccine
  efficacy.
\newblock {\em Journal of virology}, 85(2):1099--1116, 2011.

\bibitem{roy2010identification}
S.~Roy, J.~Ernst, P.~V. Kharchenko, P.~Kheradpour, N.~Negre, M.~L. Eaton, J.~M.
  Landolin, C.~A. Bristow, L.~Ma, M.~F. Lin, et~al.
\newblock Identification of functional elements and regulatory circuits by
  drosophila modencode.
\newblock {\em Science}, 330(6012):1787--1797, 2010.

\bibitem{shedden2005differential}
K.~Shedden and J.~Taylor.
\newblock Differential correlation detects complex associations between gene
  expression and clinical outcomes in lung adenocarcinomas.
\newblock In {\em Methods of Microarray Data Analysis}, pages 121--131.
  Springer, 2005.

\bibitem{storey2005significance}
J.~D. Storey, W.~Xiao, J.~T. Leek, R.~G. Tompkins, and R.~W. Davis.
\newblock Significance analysis of time course microarray experiments.
\newblock {\em Proceedings of the National Academy of Sciences of the United
  States of America}, 102(36):12837--12842, 2005.

\bibitem{straube2015linear}
J.~Straube, A.-D. Gorse, B.~E. Huang, K.-A. L{\^e}~Cao, et~al.
\newblock A linear mixed model spline framework for analysing time course
  ‘omics’ data.
\newblock {\em PloS one}, 10(8):e0134540, 2015.

\bibitem{subramanian2005gene}
A.~Subramanian, P.~Tamayo, V.~K. Mootha, S.~Mukherjee, B.~L. Ebert, M.~A.
  Gillette, A.~Paulovich, S.~L. Pomeroy, T.~R. Golub, E.~S. Lander, et~al.
\newblock Gene set enrichment analysis: a knowledge-based approach for
  interpreting genome-wide expression profiles.
\newblock {\em Proceedings of the National Academy of Sciences},
  102(43):15545--15550, 2005.

\bibitem{tsagris2014folded}
M.~Tsagris, C.~Beneki, and H.~Hassani.
\newblock On the folded normal distribution.
\newblock {\em Mathematics}, 2(1):12--28, 2014.

\bibitem{wu2012camera}
D.~Wu and G.~K. Smyth.
\newblock Camera: a competitive gene set test accounting for inter-gene
  correlation.
\newblock {\em Nucleic acids research}, 40(17):e133--e133, 2012.

\end{thebibliography}

\end{document}